S.W. Hancock[†], N. Tripathi[†], M.S. Le, A. Goffin, and H.M. Milchberg[*]

# Transverse orbital angular momentum of amplitude perturbed fields

**Abstract:** We measure the change in transverse orbital angular momentum (tOAM) per photon, $\Delta\langle L_y\rangle$, applied to an optical pulse by a pure amplitude perturbation. The results are in excellent agreement with calculations and simulations of the spatiotemporal torque based on our tOAM theory [Phys. Rev. Lett. **127**, 193901 (2021)]. The crucial factor in determining $\Delta\langle L_y\rangle$ is the spatiotemporal distribution of tOAM density in the pulse. We show that even Gaussian pulses with zero total tOAM can have net tOAM induced by an amplitude perturbation stationary in the lab frame. As a prelude to the paper, we review and clarify several recent theoretical approaches to tOAM and reemphasize several fundamental principles needed for the correct analysis of experiments and simulations.

## 1 Introduction and theoretical background

Electromagnetic (EM) fields carrying orbital angular momentum (OAM) [1] has been an intensive field of research for well over 30 years. In the most commonly studied of such fields, the OAM vector is parallel/antiparallel to the direction of propagation; examples are the Laguerre-Gauss and Bessel-Gauss beams with azimuthal phase dependence $e^{im\varphi}$ (for integer $|m|>1$) [1,2]. These longitudinal OAM beams have found a myriad of applications including optical trapping [3,4], super-resolution microscopy [5], high-harmonic generation [6], the generation of air waveguides [7], and plasma waveguides and laser-wakefield acceleration [8]. Emerging applications include free-space communications [9,10], quantum-key distribution [11], helical plasma wakes [12], and the generation of large magnetic fields in intense laser-plasma interactions [13].

Transverse OAM (tOAM), oriented orthogonal to pulse propagation, was first measured associated with spatiotemporal optical vortices (STOVs) [14], EM structures with a phase winding in spacetime [15, 16], naturally emergent in arrested nonlinear self-focusing processes such as filamentation in air [14,17] and relativistic self-guiding in plasmas [17]. Since STOVs are carried by light pulses of finite duration, they are polychromatic [14]. The understanding that STOVs were formed by spatiotemporal phase shear [14] led to the development of a 4-$f$ pulse shaper [18] specialized to linearly form STOVs by applying shear in the spatiospectral domain [19-21]. A new single-shot method, transient-grating single-shot supercontinuum spectral interferometry (TG-SSSI) [21], was used to capture the free-space propagation of STOVs in both the near and far fields [20]. Later work used a similar pulse shaper and a multi-shot scanning technique to measure STOVs in the far field only [22], and a more recent technique, resolution-limited by the pulse bandwidth, used spatially resolved spectral interferometry to characterize STOVs [23]. Since those initial studies, there has been increasing STOVs-related activity, including the demonstration of tOAM conservation in second-harmonic generation [24-27] (verifying that tOAM is carried by photons), proposed alternative methods for STOV generation [28-30], simulations of STOV-driven high-harmonic generation in gases and solids [31-33], application to laser wakefield acceleration [34,35], the generation of other structured light pulses with embedded STOVs [36-38], and the generation and measurement of spatiotemporal acoustic vortices [39,40].

In this paper, we measure the change in tOAM applied to a light pulse by a pure amplitude modulation. The results are in excellent agreement with our tOAM theory [41], and complement our recent results on pure phase perturbations [42]. Importantly, we find that the crucial factor in determining the change in tOAM is the spatiotemporal distribution of tOAM density in the pulse. This has the consequence that even Gaussian pulses with zero total tOAM can have net tOAM induced by a stationary amplitude perturbation. The experiments are also validated against numerical simulations.

Before presenting our experiments and their theoretical analysis, we first briefly review and clarify recent theoretical approaches to tOAM. In 2021, two alternative theories were presented for the tOAM of light [41,43], with one approach [43, 44] determining that the intrinsic tOAM (say along $\hat{\mathbf{y}}$) per photon of a STOV-carrying pulse propagating along $z$ in free-space has integer values similar to longitudinal OAM (along $\hat{\mathbf{z}}$) [1]. This was based on the assertion that the tOAM operator was $L_y = -i(\xi\,\partial/\partial x - x\,\partial/\partial\xi)$, identical in form to the longitudinal OAM operator $L_z = -i(x\,\partial/\partial y - y\,\partial/\partial x)$. Here, $\xi = v_g t - z$ is a lab frame coordinate local to the pulse moving at the group velocity $v_g$. This choice of $L_y$ incorrectly assumes that EM energy density vortically circulates along both the $x$ and $\xi$ axes. By contrast, the approach in [41] demonstrated from first principles that the tOAM operator is

$$L_y = (\mathbf{r}\times\hat{\mathbf{p}}_{\mathrm{ST}})_y = -i\left(\xi\frac{\partial}{\partial x} + \beta_2 x\frac{\partial}{\partial\xi}\right), \tag{1a}$$

enabling calculation of tOAM per photon $\langle L_y\rangle$ using only the electric field $E$,

$$\langle L_y\rangle = \langle E|L_y|E\rangle/\langle E|E\rangle, \tag{1b}$$

which yields half-integer values for $\langle L_y\rangle$. In Eq. (1a), $\hat{\mathbf{p}}_{\mathrm{ST}} = -i(\nabla_\perp - \beta_2\hat{\boldsymbol{\xi}}\,\partial/\partial\xi)$ is the spatiotemporal linear momentum operator, $\nabla_\perp$ is the transverse gradient, $\beta_2 = v_g^2 k_0 k_0''$ is the dimensionless group velocity dispersion, $k_0 = k(\omega_0)$ is the centre wavenumber of the field, and $k_0'' = \partial^2 k/\partial\omega^2 |_{\omega=\omega_0}$. In vacuum or very dilute or nondispersive media, $\beta_2 = 0$ and $L_y = -i\xi\,\partial/\partial x$. In Eq. (1b), $\langle E|L_y|E\rangle = \int d^3\mathbf{r}E^* L_y E$, $\langle E|E\rangle = \int d^3\mathbf{r}|E|^2$, and the integrals are over all spacetime, with $d^3\mathbf{r} = dxdyd\xi$. The half-integer values of tOAM originate from the fact that EM energy density in a STOV pulse is limited by special relativity to vertically circulate only transverse to propagation [41,42]. The operator formulation of Eqs. (1a) and (1b) gives the same half-integer tOAM per photon computed directly from the $\mathbf{E}$ and $\mathbf{H}$ fields [42,45]

$$\langle \mathbf{L}\rangle = \frac{2k_0}{U}\int d^3\mathbf{r}\,(\mathbf{r}-\mathbf{r}_c)\times\mathbf{E}\times\mathbf{H}^*, \qquad (2)$$

while the calculations in [43] do not, nor do they conserve tOAM with propagation, as shown in [42,46]. Here $U = \int d^3\mathbf{r}\,(|\mathbf{E}|^2 + |\mathbf{H}|^2)$ is proportional to the field energy, $k_0$ is the centre wavenumber of the field, and $\mathbf{r}_c = U^{-1}\int d^3\mathbf{r}\,\mathbf{r}(|\mathbf{E}|^2 + |\mathbf{H}|^2)$ is the pulse centre of energy. Note that the integrals of Eq. (1b) and Eq. (2) are taken with respect to the centre of energy (as opposed to the "photon centroid", as in [44]) so that $\langle L_y\rangle$ is the *intrinsic* (origin independent) tOAM of photons, $\langle L_y^{(i)}\rangle$.

[†]S. W. Hancock and N. Tripathi have contributed equally to this paper and should be considered as co-first authors.
[*]**Corresponding author: H. M. Milchberg**, Institute for Research in Electronics and Applied Physics, University of Maryland, College Park, USA; milch@umd.edu; https://orcid.org/0000-0003-0338-3636
**Scott W. Hancock*:** Institute for Research in Electronics and Applied Physics, University of Maryland, College Park, USA; shancoc3@umd.edu; https://orcid.org/0000-0001-5132-5516
**Nishchal Tripathi*:** Institute for Research in Electronics and Applied Physics, University of Maryland, College Park, USA; tripathi@umd.edu; https://orcid.org/0009-0004-6283-2424
**Manh S. Le:** Institute for Research in Electronics and Applied Physics, University of Maryland, College Park, USA; mle1234@umd.edu; https://orcid.org/0000-0003-4444-7894
**Andrew Goffin:** Institute for Research in Electronics and Applied Physics, University of Maryland, College Park, USA; agoffin@umd.edu; https://orcid.org/0000-0001-7406-331X

Another tOAM theory [47] was presented several years after [40,41], but as demonstrated in [42], ref. [47] produces results inconsistent with Eq. (2), incorrectly generating a non-zero value for the *extrinsic* (origin-dependent) tOAM $\langle L_y^{(e)}\rangle$ of a STOV pulse. This cancels out the intrinsic tOAM, giving a total $\langle L_y^{tot}\rangle = \langle L_y^{(e)}\rangle + \langle L_y^{(i)}\rangle = 0$, an incorrect result from which the abstract of ref. [47] asserts that applications such as STOV-induced particle rotation would be precluded. Because $\langle L_y^{(e)}\rangle$ is dependent on the choice of origin, the asserted cancelation implies that the intrinsic tOAM $\langle L_y^{(i)}\rangle$ also depends on choice of origin, an obviously erroneous conclusion.

To conclude this section, it is important to reinforce several fundamental principles at the root of these discussions and which underlie the correct analysis of experiments: (A) To calculate the intrinsic tOAM either by Eq. (1) or (2), one must use the centre of energy as the origin, so that $\langle L_y^{(e)}\rangle = 0$ and $\langle L_y^{tot}\rangle = \langle L_y^{(i)}\rangle$. (B) In cases where there is no spin-orbit coupling of light, such as for a $\hat{\mathbf{y}}$-linearly polarized paraxial beam, $\langle L_y^{tot}\rangle$ is conserved with propagation for a general choice of origin. If the origin is the centre of energy, $\langle L_y^{tot}\rangle = \langle L_y^{(i)}\rangle$ is conserved. (3) Extrinsic OAM is not a fundamental property of photons. When tOAM-carrying photons interact with matter [42], the physical interaction—such as the induced rotation of particles [47] — is independent of the origin chosen by a theorist.

## 2 tOAM and Amplitude Modulations

### 2.1 Theory

If we consider an initial pulse $A_s(x,\xi) = |A_s(x,\xi)|e^{i\phi_s(x,\xi)}$ and a spatiotemporal perturbation $\Gamma(x,\xi) = |\Gamma(x,\xi)|e^{i\Delta\phi_p(x,\xi)}$, where $\phi_s(x,\xi)$ and $\Delta\phi_p(x,\xi)$ are real functions, the change in tOAM per photon induced by the perturbation is [42]

$$\begin{aligned}\Delta\langle L_y\rangle &= \langle L_y\rangle_{\rm sp} - \langle L_y\rangle_{\rm s}\\ &= iu_{\rm sp}^{-1}\int dxd\xi\,\Big[|A_s|^2|\Gamma|^2 L_y\Delta\phi_p\\ &\qquad + |A_s|^2\left(|\Gamma|^2 - \frac{u_{\rm sp}}{u_s}\right)L_y\phi_s\Big]. \qquad (3)\end{aligned}$$

Here $A_{\rm sp} = \Gamma(x,\xi)A_s(x,\xi)$ is the perturbed pulse, $u_s = \int dxd\xi\,|A_s(x,\xi)|^2$, and $u_{\rm sp} = \int dxd\xi\,|A_s(x,\xi)|^2|\Gamma(x,\xi)|^2$. In this paper we consider pure amplitude perturbations ($\Delta\phi_p(x,\xi) = 0$) so that Eq. (3) becomes

$$\Delta\langle L_y\rangle = iu_{\rm sp}^{-1}\int dxd\xi\,\left(|\Gamma|^2 - \frac{u_{\rm sp}}{u_s}\right)|A_s|^2 L_y\phi_s. \qquad (4)$$

This equation shows that for a pure amplitude perturbation to change the tOAM of a pulse, that pulse must initially have nonzero tOAM density $M_y(x,\xi) = A_s^* L_y A_s = i|A_s|^2 L_y\phi_s$, or $L_y\phi_s \neq 0$.

We start with a STOV pulse of integer topological charge $l$,

$$A_{\rm STOV}(x,\xi) = \left(\frac{\xi}{w_{0\xi}} \pm i\frac{x}{w_{0x}}\right)^{|l|} A_G(x,\xi), \qquad (6)$$

where $A_G(x,\xi) = A_0 e^{-x^2/w_{0x}^2 - \xi^2/w_{0\xi}^2}$ is a Gaussian envelope, $w_{0x}$ and $w_{0\xi}$ are space-like and time-like scale widths of the pulse, and for later use we define the spacetime asymmetry ratio $\alpha = w_{0\xi}/w_{0x}$. The form of the STOV pulse in Eq. (6) applies near the beam waist $z = 0$ for $z/z_R \ll 1$, where $z_R = k_0 w_{0x}^2/2$ is the Rayleigh length; evolution of the pulse outside of this region is governed by the linear theory presented in [41]. We apply to this pulse a stationary pure amplitude perturbation of the form

$$\Gamma(x,\xi) = 1 - \exp(-[(x-x_0)/h_x]^{20}), \qquad (7)$$

transversely centred at $x = x_0$, and which removes pulse energy over a transverse spatial scale $\sim 2h_x$. Figures 1 (a), (b), and (c) plot the intensity envelope $|A_{\rm STOV}(x,\xi)|^2$, phase $\Phi(x,\xi) = \arg(A_{\rm STOV})$, and tOAM density $M_y(x,\xi)$ of the incident STOV pulse. In the frames, the pulse propagates from right to left. The pattern for $M_y$ in Fig. 1(c) originates from $iL_y\phi_s = \xi\,\partial\phi_s/\partial x$ contributing mainly near the spatiotemporal phase locations $\Phi$=0 and $\Phi = \pi$. Figure 1(d) plots $\Delta\langle L_y\rangle$ induced by the obstruction (Eq. (7)) as a function of $x_0/w_{0x}$ and $h_x/w_{0x}$, computed using Eq. (4). It is seen that the transverse range $x_0/w_{0x}$ over which the perturbation induces a negative $\Delta\langle L_y\rangle$ increases with $h_x/w_{0x}$. This stems from the increasing blockage of the high $M_y$ regions concentrated near $x = 0$ (see Fig. 1(c)) but without proportionally removing pulse energy; this reduces the tOAM per photon. Interestingly, an obstruction can also induce positive changes in $\Delta\langle L_y\rangle$, as seen at the yellow edges of the angular opening in Fig. 1(d). This will be further discussed below.

An important insight from Eq. (4) is that nonzero $\Delta\langle L_y\rangle$ induced by an obstruction requires that the initial pulse contain nonzero tOAM density $M_y$; total pulse tOAM, however, can be zero. As an example of this, Figs. 1(e)-(g) plot $|A(x,\xi)|^2$, $\Phi(x,\xi)$, and $M_y(x,\xi)$ of a diverging Gaussian pulse at $z = 2z_R$ and Figs. 1(i)-(k) plot the same functions at $z = 200z_R$. The nonzero $M_y(x,\xi)$ distributions originate from the curved phase fronts. Selective obstruction of tOAM density can then produce nonzero $\Delta\langle L_y\rangle$, as discussed above for the STOV pulse. For the wire placed at $z = 2z_R$ and $z = 200z_R$, Figs. 1(h) and 1(l) plot $\Delta\langle L_y\rangle$ as a function of $x_0/w_{0x}$ and $h_x/w_{0x}$. It is seen that changes in tOAM at $2z_R$ are much smaller than at $200z_R$. They are also smaller than our measurement error, as discussed in Sec. 2.3.

Nevertheless, it is worth discussing the physical interpretation of Fig. 1(h). As $h_x/w_{0x}$ increases, there is a narrowing of the $x_0$ range (centred on $x_0 = 0$) for which $\Delta\langle L_y\rangle = 0$. For $h_x/w_{0x} > \sim 1$, positive and negative regions of $\Delta\langle L_y\rangle$ appear in the positive and negative wings of $x_0/w_{0x}$. The region of $\Delta\langle L_y\rangle = 0$ centred on $x_0 = 0$ is explained by the symmetric blockage of tOAM density above and below the $z$ (or $\xi$) axis; this symmetry ensures that the integral of Eq. (4) is zero. Because the beam is diverging, when $x_0 > 0$ the obstruction blocks more negative than positive $M_y$. The integral of Eq. (4), having its negative contributions reduced, gives $\Delta\langle L_y\rangle > 0$. And when $x_0 < 0$, the positive tOAM density contributions in Eq. (4) are reduced, giving $\Delta\langle L_y\rangle < 0$. For larger $h_x/w_{0x}$, there is greater sensitivity to asymmetric blockage and the $\Delta\langle L_y\rangle = 0$ zone narrows as described above. If, however, the wire had been placed at the Gaussian beam waist, we would have $\Delta\langle L_y\rangle = 0$ for all $x_0/w_{0x}$ and $h_x/w_{0x}$ because $M_y = 0$ everywhere at the waist.

In Fig. 1(l) ($z = 200z_R$), the greater tOAM density and extent contributes to larger $\Delta\langle L_y\rangle$, with similar physical arguments as in Fig. 1(h) for the dependence of $\Delta\langle L_y\rangle$ on $x_0/w_{0x}$ and $h_x/w_{0x}$.

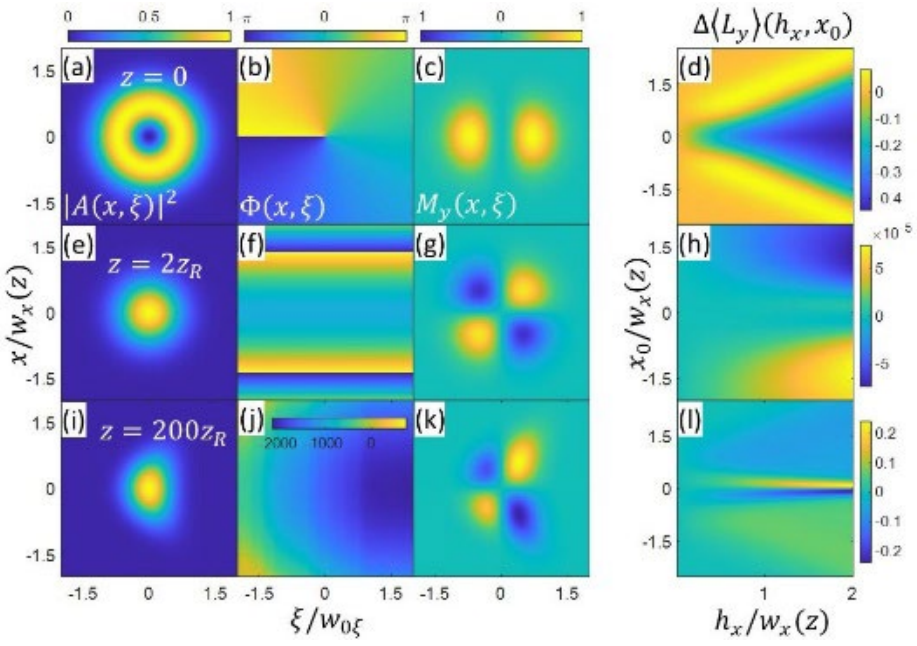

**Fig. 1**: **Unperturbed $\boldsymbol{l = 1}$ STOV pulse:** (a) intensity envelope $|A(x,\xi)|^2$, (b) spatiotemporal phase $\Phi(x,\xi)$, and (c) tOAM density $M_y(x,\xi)$. (d) change in tOAM per photon $\Delta\langle L_y\rangle$ as a function of wire width and vertical position $x_0$. Wire is at beam waist $z = 0$. **Unperturbed Gaussian pulse at $\boldsymbol{z = 2z_R}$:** (e) intensity envelope $|A(x,\xi)|^2$, (f) spatiotemporal phase $\Phi(x,\xi)$, and (g) tOAM density $M_y(x,\xi)$. (h) change in tOAM per photon $\Delta\langle L_y\rangle$ as a function of wire width and $x$-position. Wire is at $z = 2z_R$. **Unperturbed Gaussian pulse at $\boldsymbol{z = 200z_R}$:** (i) intensity envelope $|A(x,\xi)|^2$, (j) spatiotemporal phase $\Phi(x,\xi)$ (here with full phase scale), and (k) tOAM density $M_y(x,\xi)$. (l) change in tOAM per photon $\Delta\langle L_y\rangle$ as a function of wire width and $x$-position. Wire is at $z = 200z_R$.

### 2.2 Experimental setup

The stationary obstruction modeled in the prior section is a 50 μm diameter tungsten wire placed at the beam waist $z = 0$, as shown in Fig. 2. The incident pulse is either a Gaussian pulse or a $l = 1$ STOV pulse, with the latter depicted in the figure. The pulse emerges from a $4f$ pulse shaper, propagates through air to interact with the wire, and then is measured by TG-SSSI [21]. The recovered spatiotemporal field is then analyzed to compute $\langle L_y\rangle$ and $\Delta\langle L_y\rangle$ using Eq. (1b) and Eq. (4).

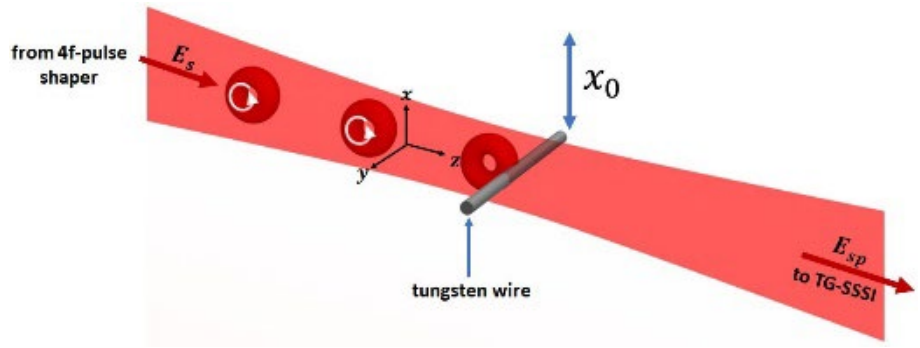

**Fig. 2:** Interaction region of the experiment. The focused STOV or Gaussian pulse $E_s$ is partially obstructed by a vertically translatable 50 μm diameter tungsten wire. The resulting perturbed beam is then collected by the TG-SSSI diagnostic (see text).

In more detail, the output of a 1-kHz repetition-rate Ti:Sapphire laser was split into 4 pulses: (1) input to the $4f$-pulse shaper yielding an output pulse $E_s$ (18 μJ (Gaussian), 26 μJ (STOV)), (2) input into a bandpass filter for a narrower bandwidth spatial interferometry reference pulse $\mathcal{E}_i$ for TG-SSSI and (3) twin probe and reference supercontinuum (SC) pulses $E_{pr}$ and $E_{ref}$ (with bandwidth $\Delta\lambda_{SC} \approx 330\text{nm}$ centred at $\lambda_{SC} \approx 555\text{nm}$. Similar experimental setups using TG-SSSI are shown in refs. [20, 42], accompanied by explanations for all of the needed beams.

The pulse $E_s$ was focused using an 85 cm focal length lens to its waist, where the tungsten wire could be moved transverse to the beam path by a vertical stage, thereby adjusting $x_0$. The waist region was imaged using all-reflective optics, delivering the perturbed pulse $E_{sp}$ into a 100 μm thick BK7 "witness plate". There, it was overlapped spatiotemporally with $\mathcal{E}_i$ to form a transient grating, which was probed collinearly and temporally with the SC probe pulse $E_{pr}$. In advance of these 3 pulses, a SC reference pulse $E_{ref}$ propagated though the witness plate collinear to $E_{pr}$. The output face of the witness plate was imaged onto the slit of an imaging spectrometer and spectral interferograms were collected using a CMOS camera. $E_s$ and $\mathcal{E}_i$ were filtered from the imaging system after the witness plate with a dielectric mirror. An adjustable slit near the Fourier plane of the $4f$-pulse shaper was used to decrease the bandwidth of the pump pulse $E_s$, thereby increasing $w_{0\xi}$ as well as $\alpha$. Additionally, an adjustable iris was placed just before the pump focusing lens to increase $w_{0x}$. This was necessary to keep the scale of Gaussian pulse larger than the tungsten wire.

### 2.3 Results and Discussion

Figures 3(a) and (b) show the TG-SSSI-extracted experimental results and simulations, respectively, for Gaussian pulses. The rows of panel (a) show the measured intensity envelope $|A(x,\xi)|^2$, spatiotemporal phase $\Phi(x,\xi)$, and tOAM density $M_y(x,\xi)$. Column (i) plots the unperturbed Gaussian pulse with $\alpha = w_{0\xi}/w_{0x} = 29\mu\text{m}/159\mu\text{m} \sim 0.18$. For air, $\beta_2 \cong 1.5 \times 10^{-5}$ is very small and taken to be zero for our computations of $\langle L_y\rangle$ and $\Delta\langle L_y\rangle$ from the extracted fields using Eqs. (1b) and Eq. (4). Results for 3 positions of the wire ($x_0 \approx -50, 0, 50$ μm) with respect to the unperturbed pulse centre of energy are shown in columns (ii), (iii) and (iv). In the top row, red dots mark the centre of energy. The wire was placed longitudinally at $z = 0$ and $z = 2z_R$; both results were the same within measurement error, so we show only the $z = 0$ case.

In column (i), weak modulations are seen in the intensity and the tOAM density from diffraction off the iris before the pump focusing lens. The magnitude of the tOAM density $M_y$ is very low; upon integration, $\langle L_y\rangle \sim 0.01 \rightarrow 0.0$ to one significant figure, which lies inside the measurement error of $\sim \pm 0.01$. Note that $M_y \sim 0$ right at the centre of energy at $z = 0$, since the phase front there is essentially flat. Passing through the centre of energy for increasing $\xi$, $M_y$ flips sign from negative to positive consistent with the sign change imposed by running the $\xi$-axis through the centre of energy.

Figures 3(a)(ii-iv) show the measured results for the 3 $x$-positions of the wire. The vertical modulations in $|A(x,\xi)|^2$ are the diffractive effect of the wire edges on $E_s$. Owing to conservation of OAM, imaging anywhere after the wire will give the same OAM result. The extracted phase $\Phi(x,\xi)$ shows a horizontal strip with no signal that tracks the location of the wire. This is due to the lack of signal in that region caused by the obstruction. However, this does not affect the computed results since the field amplitude in that region is zero. The bottom row of Fig. 3(a) shows the same qualitative pattern of very low

$M_y(x,\xi)$, with $\Delta\langle L_y\rangle \sim 0.0$ for all $x$-positions of the wire, consistent with the prediction in Fig. 1(h) of $\Delta\langle L_y\rangle$ much smaller than the measurement error.

Finally, Fig. 3(b) plots simulated (i) unperturbed and (ii-iv) perturbed Gaussian intensity envelopes $|A(x,\xi)|^2$ with corresponding phase $\Phi(x,\xi)$ included as an inset in Fig. 3(b)(i). The wire is placed at $z=0$. The tOAM density is zero everywhere and is not plotted; $\Delta\langle L_y\rangle = 0$ is in good agreement with the results of Fig. 3(a). We note that space constraints in the experimental setup precluded placing a wire more than $2z_R - 3z_R$ from the beam waist; longer distances where $\Delta\langle L_y\rangle$ was above noise could only be explored by simulations, as in Fig. 1(i)-(l). However, the agreement between theory and experiment for STOVs (demonstrated below in Fig. 3(c)-(d)) lends confidence to our simulations of wire-perturbed Gaussian beams.

Figure 3(c) shows the experimental results for wire obstruction of an $l=1$ STOV pulse with $\alpha = w_{0\xi}/w_{0x} = 39\mu m/84\mu m = 0.46$. The rows plot the measured pulse intensity envelope $|A(x,\xi)|^2$, its spatiotemporal phase $\Phi(x,\xi)$, and the tOAM density $M_y(x,\xi)$. Again, the modulations in $|A(x,\xi)|^2$ are caused by the diffractive effect of the wire edges on $E_s$. These modulations are tilted, unlike those in Fig. 3(a) (top row), because the phase fronts within a STOV pulse are tilted due to edge dislocation [14]. The extracted phase $\Phi(x,\xi)$ shows a similar horizontal strip as in Fig. 3(a) (second row) due to the lack of signal in the wire-obstructed region. The centres of energy are marked in the top row with red dots. Figure 3(c)(i) plots the measured intensity, phase, and tOAM density of the unperturbed STOV pulse, with columns (ii)-(v) plotting results for wire positions $x_0 = 0, -20, -40, -60$ μm. The bottom row of panel (c) shows that $M_y(x,\xi)$ is positive everywhere, as in Fig. 1(c), stemming from the fact that $iL_y\phi_s = \xi\,\partial\phi_s/\partial x$ (see Eq. (4)) does not change sign across the temporal centre of energy. Note that $M_y(x,\xi)$ for the STOV pulse is much greater than for the Gaussian pulse. Centering the obstruction on the STOV pulse ($x_0 = 0$ μm, column (ii)) yields the largest magnitude change in tOAM, $\Delta\langle L_y\rangle = -0.06$, because the obstruction blocks the regions of highest $M_y(x,\xi)$. As the wire is shifted away from the centre of energy, the magnitude of $\Delta\langle L_y\rangle$ first drops but then increases, as predicted by Fig. 1. This is because $M_y(x,\xi)$ has a smaller spatial extent than $|A(x,\xi)|^2$; as the wire moves away from the unperturbed STOV centre of energy, it blocks more energy density than $M_y(x,\xi)$, yielding an increase in the tOAM per photon $\Delta\langle L_y\rangle$. This is shown quantitatively in the third row of Fig. 3(c)(i) by fitting a Gaussian to $M_y(x,\xi)$ at the $\xi$ location of the vertical green dashed line, giving a FWHM of 104 μm. A similar fit to $|A(x,\xi)|^2$ at the same $\xi$ location gives a FWHM of 128 μm. Thus, when the wire is located at $x_0 = -60$ μm, it obstructs more energy density than tOAM density.

The top row of Fig. 3(d) plots simulations of $|A(x,\xi)|^2$ for (i) unperturbed and (ii-v) wire obstruction at $x_0 = 0, -20, -40, -60$ μm. The inset in (i) shows $\Phi(x,\xi)$ for (i-v). The bottom row plots the corresponding $M_y(x,\xi)$, and shows the computed $\langle L_y\rangle$. Figure 3(e) overlays the simulated and experimental values for $\Delta\langle L_y\rangle$, where additional experimental points are included. Agreement is excellent. The simulation corresponds to that of Fig. 1, using $h_x = 0.3w_{0x}$, and rescaling it from the case of $\alpha = 1$ to that of the experimental case of $\alpha = 0.46$. Plots of the additional experimental data points for intensity envelope, phase and tOAM density can be found in the supplementary material [48].

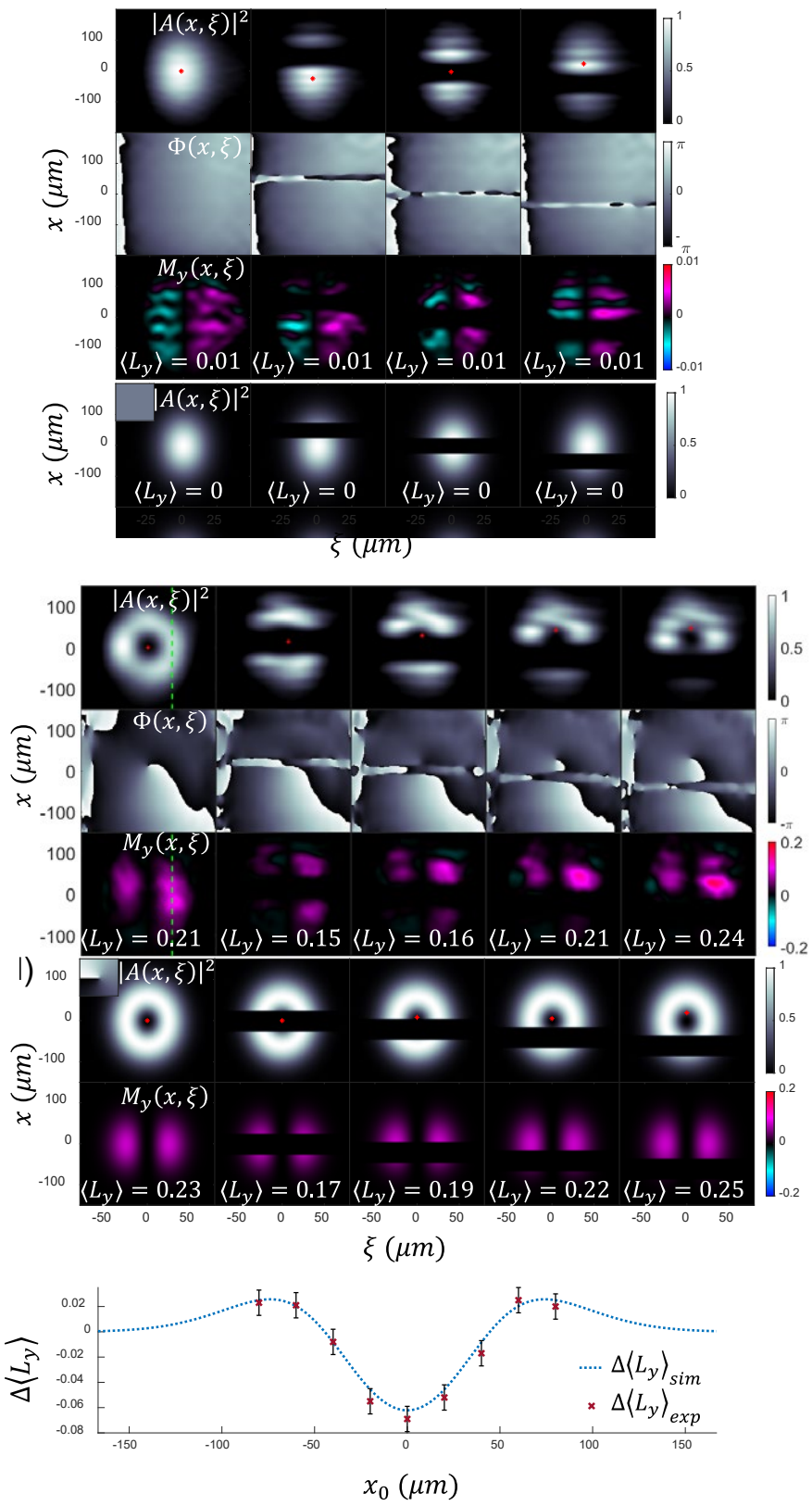

**Fig. 3:** Experimental results and simulations for 50 μm diameter tungsten wire perturbation of a Gaussian pulse or an $l=1$ STOV pulse. **(a) Gaussian pulse**: rows, top to bottom, plot $|A(x,\xi)|^2$, $\Phi(x,\xi)$, and $M_y(x,\xi)$. Column (i) is unperturbed Gaussian, and columns (ii)-(iv) show results for the wire centred at $x_0 = -50, 0, 50$ μm with respect to the unperturbed pulse centre of energy. **(b)** Simulation corresponding to (a). **(c) STOV pulse** ($l=1$): rows, top to bottom, plot $|A(x,\xi)|^2$, $\Phi(x,\xi)$, and $M_y(x,\xi)$. Column (i) is unperturbed STOV, and columns (ii)-(v) show results for the wire centred at $x_0 = 0, -20, -40, -60$ μm. The red dots mark the centres of energy and the measured tOAM is shown on the panels. (d) Simulation corresponding to (c). The insets in (b,d) show the corresponding phases. (e) Overlay of the simulated and experimental values for $\Delta\langle L_y\rangle$ from (c).

In summary, we have experimentally and theoretically demonstrated that a pure amplitude perturbation to an optical pulse can change its transverse OAM (tOAM) per photon, provided that the initial pulse contains nonzero tOAM density. The change in tOAM per photon—either positive or negative— depends on where in the pulse the perturbation acts relative to the tOAM density distribution. This effect means that even a pulse with zero total tOAM but nonzero tOAM density-- such as in a converging or diverging Gaussian beam-- can have net tOAM induced on it by a carefully placed amplitude perturbation. Finally, we presented a brief review of recent theories of tOAM in order to reemphasize several fundamental principles needed for the correct analysis of experiments and simulations.

**Research funding:** This work is supported by the Air Force Office of Scientific Research (Grant No. FA9550-21-1-0405) and the Office of Naval Research (Grants No. N00014-17-1-2705 and No. N00014-20-1-2233)

**Author contribution:** All authors have accepted responsibility for the entire content of this manuscript and approved its submission.

**Conflict of interest**: Authors state no conflict of interest.

**Data availability statement**: The datasets generated and/or analysed during the current study are available from the corresponding author upon reasonable request.